\magnification=\magstep1
\overfullrule=0pt
\baselineskip=11pt
\def\BZT{{\rm Z{\hbox to 3pt{\hss\rm Z}}}}
\def\BZS{{\hbox{\sevenrm Z{\hbox to 2.3pt{\hss\sevenrm Z}}}}}
\def\BZSS{{\hbox{\fiverm Z{\hbox to 1.8pt{\hss\fiverm Z}}}}}
\def\BZ{{\mathchoice{\BZT}{\BZT}{\BZS}{\BZSS}}}
\def\BCT{\,\hbox{\hbox to -3pt{\vrule height 6.5pt width .2pt\hss}\rm C}}
\def\BCS{\,\hbox{\hbox to -2.2pt{\vrule height 4.5pt width .2pt\hss}$
    \scriptstyle\rm C$}}
\def\BCSS{\,\hbox{\hbox to -2pt{\vrule height 3.3pt width
    .2pt\hss}$\scriptscriptstyle \rm C$}}
\def\BC{{\mathchoice{\BCT}{\BCT}{\BCS}{\BCSS}}}
\font\ti=cmbx10 scaled \magstep1
\def\q#1{\lbrack #1\rbrack}
\def\bibitem#1{\parindent=8mm\item{\hbox to 6 mm{\q{#1}\hfill}}}
\rightline{hep-th/9608121}
\rightline{BONN-TH-96-05}
\rightline{August 1996}
\bigskip
\centerline{\ti On the Seiberg--Witten approach to electric--magnetic duality}
\medskip
\centerline{Werner Nahm}
\centerline{Physikalisches Institut der Universit\"at Bonn,
Nu{\ss}allee 12, D--53115 Bonn}
\centerline{e-mail: werner@avzw01.physik.uni-bonn.de}
\bigskip
{\bf Abstract:} {\it Electric-magnetic duality allows to calculate the
central charges of $N=2$ supersymmetric theories with massless hypermultiplets
as derivatives of simple modular forms. The procedure reproduces the 
Seiberg-Witten results for $N_f=0,2,3$ in a uniform way, but indicates open
problems for $N_f=1$.} 
\bigskip
Moduli spaces of $N=2$ supersymmetric quantum field theories are complex
varieties. Moreover, the central charge of the supersymmetry depends
analytically on the parameters. When the moduli spaces can be compactified,
this allows an exact determination of the central charges. For gauge theories
in four dimensions, the hypothesis of electric-magnetic duality [1-4] implies
compactness. More precisely, the moduli space can be compactified by adding a
few points with neighbourhoods described by perturbation theory. Using these
ideas, Seiberg and Witten determined the central charges of the $SU(2)$ gauge
theories with massless hypermultiplets [5,6]. The case of a single
hypermultiplet in the adjoint representation is simple and was understood
before, and the case of $N_f=4$ massless hypermultiplets in the fundamental
representation is rather analogous. The treatment of $N_f=0,1,2,3$ such
hypermultiplets, however, constituted a breakthrough in quantum field theory.
Nevertheless, the uniqueness of the solutions given by Seiberg and Witten was
hard to control, in part because their use of elliptic curves appears somewhat
ad hoc. 

Here we shall consider the problem in a way which is closer in spirit to
S-matrix theory. Under physically very plausible assumptions, the latter
produces two dimensional scattering matrices which are unique up to CDD poles.
The selection of the correct matrices is made by simplicity assumptions and
perturbation theory.

\smallskip
Consider $SU(2)$ gauge theories with gauge symmetry broken to $U(1)$.
For the moment, we only consider the massless case, but a generalization to
massive hypermultiplets should be easy. In this case, the only continuous
parameters are given by the charge lattice. The electric and magnetic charges
of the various superselection sectors of the theory form a group $\Sigma$ 
isomorphic to $\BZ^2$. The Coulomb force yields a quadratic form on $\Sigma$,
such that this group can be embedded in the complex plane $\BC$ with its
standard metric. The area of the fundamental domain $\BC/\Sigma$ is fixed by
Dirac's quantization condition. When an embedding is chosen, the real part of
any $\sigma\in\Sigma$ may be called electric charge and the imaginary part
magnetic charge. Electric-magnetic duality states that the choice of the
embedding has no physical meaning, however. In nature, the smallest charges
are called electric, in other words, the shortest non-vanishing
$\sigma\in\Sigma$ is chosen to be real, but when the parameters are varied,
this convention makes no global sense.

For small electric charge and parity conservation, let $2\sigma_2$ be the
charge of a massive gauge boson and $\sigma_1$ the charge of a magnetic
monopole. We put $\tau=\sigma_1/\sigma_2$ and choose the signs such that
$\tau$ varies over the upper complex half plane ${\cal H}$. When there are
no fermions of charge $\sigma_2$, then $\tilde\tau=\tau/2$ is another natural
choice.

A base change $(\sigma_1,\sigma_2)\mapsto
(A\sigma_1+B\sigma_2,C\sigma_1+D\sigma_2)$, ${AB\choose CD}\in SL(2,\BZ)$,
yields an action $\tau\mapsto (A\tau+B)/(C\tau+D)$ of the modular group 
$\Gamma$. Thus charge lattices are parametrized by points in
${\cal H}/\Gamma$. For theories without further continuous parameters,
electric-magnetic duality states that the parameter space of the theory
is given by a finite covering of ${\cal H}/\Gamma$.

Often, an action of a discrete symmetry group is defined for all parameter 
values of a theory. With respect to the action of such a group, $\Sigma$
decomposes into several cosets. Only those elements of the modular group
which preserve the cosets yield reparametrizations of a theory. In the $SU(2)$
theories with $N_f$ hypermultiplets, the discrete symmetry group depends on
$N_f$. For $N_f=0$, states with charges $m\sigma_1+(2n+1)\sigma_2$ do not
exist, such that only modular transformations with even $B$ are allowed.
These form the subgroup $\Gamma^0(2)$ of the modular group. Equivalently,
these transformations of $\tau$ can be considered as modular transformations
of $\tilde\tau$ with even $C$, such that the parameter space of the theory
is ${\cal H}/\Gamma^0(2)$ in terms of $\tau$, and
${\cal H}/\Gamma_0(2)$ in terms of $\tilde\tau$.

For $N_f>0$, the global symmetry group is $Spin(2N_f)$ and the various points
of the charge lattice yield scalar, vector, or half-spinor representations of
this group. In particular, one has a homomorphism $\rho$ from $\Sigma$ into
the center of $Spin(2N_f)$. It is sufficient to know its values on generators
$m\sigma_1+n\sigma_2$ of $\Sigma$, for which $m,n$ have no common prime
factor. For $N_f=2$, no generator of $\Sigma$ yields a scalar. The generators
$(2m+1)\sigma_1+n\sigma_2$ yield half-spinors and have to transform into each
other. Thus only modular transformations with even $C$ are allowed. These form
the subgroup $\Gamma_0(2)$ of the modular group. All $\Gamma_0(2)$ make
physical sense, since the two different half-spinors are related by an outer
automorphism. For $N_f=4$, the two half-spinors and the vector can all be
permuted by outer automorphisms, such that the full modular group yields
possible reparametrizations. For $N_f=3$, generators of the form
$4m\sigma_1+n\sigma_2$ are vectors, those of the form
$(4m+1)\sigma_1+n\sigma_2$ and $(4m+3)\sigma_1+n\sigma_2$ are
half-spinors, those of the form $(4m+2)\sigma_1+n\sigma_2$ are scalars.
Thus only modular transformations with $C$ divisible by 4 are allowed.
These form the subgroup $\Gamma_0(4)$ of the modular group.
The case $N_f=1$ will be discussed below.

When one of the generators of $\Sigma$ becomes very short, one reaches a
perturbative regime, and a cusp of the parameter space. The perturbative
behaviour depends on the corresponding representation of $Spin(2N_f)$.
For $N_f=0,2$ there are two cusps, for $N_f=3$ there are three, and for
$N_f=4$ a single one. In detail, the fundamental domain of $\Gamma$ can be
compactified by a single cusp point given by $\tau_0=i\infty$. For
${\cal H}/\Gamma_0(2)$ there is a second inequivalent cusp at $\tau_1=0$ and
for ${\cal H}/\Gamma_0(4)$ a third one at $\tau_2=1/2$. The fundamental
domain of $\Gamma_0(4)$ is smooth, whereas its $\BZ_2$ quotient
${\cal H}/\Gamma_0(2)$ has an orbifold point at $\tau=(1+i)/2$.
Finally, the fundamental domain of $\Gamma$ has a further threefold
orbifold point at $\tau=exp(\pi i/3)$.

\smallskip
The central charge $Z$ of our supersymmetric theories is a linear function
of $\Sigma$, defined up to an anomalous phase in a finite cyclic group.
Following Seiberg and Witten, let us put $Z(\sigma_1)=a_D$, $Z(\sigma_2)=a$.
For $N_f=0$, put $\tilde a =2a$.

For a base change ${AB\choose CD}\in SL(2,\BZ)$, i.e. $\tau\rightarrow
(A\tau+B)/(C\tau+D)$, one has $a_D \rightarrow Aa_D+Ba$, $a_D\rightarrow
Ca_D+Da$. The differentials $da_D,da$ transform in the same way.

One assumes that the $a,a_D$ are holomorphic in the interior of the
moduli space. Together with the perturbative behaviour at the cusps this
implies that $c=a_D-\tau a$ is a modular form of weight $-1$.
For $N_f=4$, there is no cusp singularity either, such that $c=0$.

For $N_f=2,3$ let $c^{2(4-N_f)}\sim exp(2\pi in_k\tau')$ at the cusps
$\tau_k$, where $\tau'=(A\tau+B)/(C\tau+D)$ near $\tau=-D/C$.
Analogously, let $c^4\sim exp(2\pi in_k\tilde\tau')$
for $N_f=0$. As noted in [6], the latter case becomes isomorphic to the
$N_f=2$ case and needs not be treated separately. Thus we now restrict
ourselves to $N_f=2,3$. For $N_f=2$, the moduli space is
${\cal H}/\Gamma_0(2)$, which implies $n_2=n_0$.

Since $c^{2(4-N_f)}$ has weight $-2(4-N_f)$ one has the pole counting formula
$$n_0+4n_1+n_2=(4-N_f)(1+2n(0))\ ,$$
where $n(0)$ is the number of zeros of $c$ in the interior of
${\cal H}/\Gamma_0(4)$. The moduli space ${\cal H}/\Gamma_0(2)$ can be
written as a $\BZ_2$ quotient of ${\cal H}/\Gamma_0(4)$. Thus we can use the
same formula, with the $\BZ_2$ symmetry condition $n_2=n_0=1$.
Positive values of the $n_k$ correspond to asymptotic freedom of the
corresponding perturbative theories. In particular, $n_0=1$.
This is just sufficient to saturate the pole counting formula
with the minimal values $n(0)=0$, $n_1=0$. For $N_f=2$ one has
$n_2=n_0=1$, whereas $N_f=3$ yields $n_2=0$, too. Solutions with more zeros
would have further asymptotically free domains, for which there is no
indication. In this sense, the Seiberg-Witten solution is unique.

The poles determine $c$ up to a multiplicative constant,
which just sets the length scale. Indeed, the linear combinations of two
independent functions would have a continuously variable zero.
Moreover, all such functions are simple products of $\eta(\tau)$,
$\eta(2\tau)$ and $\eta(4\tau)$. The latter function is not modular on
${\cal H}/\Gamma_0(2)$, thus it is only needed for $N_f=3$. One obtains
immediately
$$c(\tau)\sim \eta(\tau)^2/\eta(2\tau)^4$$
for $N_f=2$ and analogously
$$c(\tau)\sim \eta(\tau/2)^2/\eta(\tau)^4$$
for $N_f=0$. For $N_f=3$ one finds
\vskip-5pt
$$c(\tau)\sim \eta(2\tau)^2/\eta(4\tau)^4\ .$$
In the latter case, the only contributing instanton numbers are
multiples of 4, whereas for $N_f=2$ one gets the even numbers. For
$N_f=0$, there is no such restriction.

Since $c$ has no zeros, $(da_D-\tau da)/c$ is a holomorphic differential on
${\cal H}/\Gamma_0(4)$. The latter space has the holomorphic structure of a
sphere, such that $da_D-\tau da=0$. Consequently, $a=dc/d\tau$, which yields
elementary explicit formulas for $a,a_D$. The results can be checked close to
$\tau_1,\tau_2$, where the $\beta$-function can be read off from the first
non-constant term in the $q$-expansion of $c$. The results agree with the
expected ones from supersymmetric abelian gauge theories.

\smallskip
For $N_f=1$, the procedure described above yields no consistent results when
the moduli space is assumed to be some quotient of ${\cal H}/\Gamma_0(4)$.
Instead, Seiberg and Witten have proposed another solution.
They parametrize their functions by a parameter $u$ and use an
elliptic curve $y^2=x^3-x^2u+1$. For $N_f=0,2,3$, the function $u$
is modular for $\Gamma_0(4)$. For $N_f=1$, however, the relation between
$u$ and the modular parameter $\tau$ is given by
$${u^6\over 4u^3-27}=2^{-8}j(\tau)\ .$$
The value $u=0$ corresponds to the orbifold point $\tau^o=exp(\pi i/3)$,
in the neighbourhood of which one finds $u^2\sim \tau-\tau^o$.
Thus $u$ has a square root cut at $\tau^o$. Thus the parameter space of
the theory becomes a ramified double cover of ${\cal H}/\Gamma$.
This conjecture is most probably correct, since it agrees with the
results of two-instanton calculations [7,8].

If this form of the parameter space is accepted, it again is easy to find
$c$. Let $c_{\pm}$ describe the two branches of $c$. Then
$c_+^6+c_-^6$ must be a modular form of weight $-6$ with a single pole at
the cusp, thus equal to a multiple of $E_6(\tau)\eta(\tau)^{-24}$.
We assume that only one branch yields an asymptotically free domain.
Thus $c_+^6c_-^6$ must be a modular form of weight $-12$ with a single
pole at the cusp, thus equal to a multiple of $\eta(\tau)^{-24}$.
The ramification point can be moved by the choice of proportionality
constants. For a cut at $\tau^o$, we obtain
$$c_{\pm}^6\sim (E_6\pm E_4^{3/2})\eta(\tau)^{-24}\ .$$

At the orbifold point, the theory gets an unexpected $\BZ_3$-symmetry. The
charge lattice is incompatible with an $SU(3)$ gauge group, such that the
spin one gauge bosons cannot exist as mutually local fields. Indeed, the
ration $a_D/a$ is real at the orbifold point, such that it belongs to the
boundary of the stability domain of the gauge bosons.
Obviously, this case needs further investigation.

\smallskip
In the perturbative domains, the function $u$ describes the vacuum
expectation value of the square of the Higgs field. This is a function
on moduli space, thus it has weight zero. At the cusps, it is proportional to
$c^2$. This implies that it has a single zero in the interior of
${\cal H}/\Gamma_0(4)$. For $N_f=1,2$, the zero must lie at the orbifold
point of ${\cal H}/\Gamma_0(2)$, such that $u$ is fixed uniquely. For $N_f=3$,
the position of the zero is arbitrary, such that $u$ only can be determined
up to an additive constant. This makes good physical sense, since in this
case the square of the Higgs field mixes with the constant operator.

One obtains
$$u=\left({\eta(\tau)^8\over\eta(2\tau)^8}\right)+8$$
for $N_f=2$. The latter formula applies to $N_f=0$, too, when $\tau$ is
replaced by $\tilde\tau$. For $N_f=3$, ref.\ [6] puts the zero of $u$ at the
cusp $\tau_1$, which yields
\vskip-4pt
$$u={\eta(\tau)^8\over\eta(4\tau)^8}\ .$$
\vskip-2pt

For $N_f=1$, the double cover of Seiberg and Witten yields
$$u^3_\pm=2^{-7}(E_4(\tau)^3\pm E_6(\tau)E_4(\tau)^{3/2})\eta(\tau)^{-24}\ .$$
In all cases, the multiplicative normalization just amounts to a choice of
scale.

Finally, let us remark that $da/du$ has weight $1$ and behaves at the
cusps like $c^{-1}$. When none of these forms has poles in the interior of
moduli space, this implies $da/du\sim c^{-1}$. 

\bigskip
{\sl Acknowledgement:} I would like to thank E.\ Witten for critical remarks.
\bigskip
\line{\bf References\hfil}
\smallskip
\bibitem{1} P.\ Goddard, J.\ Nuyts and D.\ Olive,
 Nucl.\ Phys.\ {\bf B125} (1977) 1
\bibitem{2} C.\ Montonen and D.\ Olive, Phys.\ Lett.\ {\bf 72B} (1977) 117
\bibitem{3} E.\ Witten and D.\ Olive, Phys.\ Lett.\ {\bf 78B} (1978) 97
\bibitem{4} D.\ Olive in: Monopoles in Quantum Field Theory, N.\ Craigie
 et al.\ eds., Proc.\ Trieste 1981, World Scientific, Singapore 1982
\bibitem{5} N.\ Seiberg and E.\ Witten, Nucl.\ Phys.\ {\bf B426} (1994) 19
\bibitem{6} N.\ Seiberg and E.\ Witten, Nucl.\ Phys.\ {\bf B431} (1994) 484
\bibitem{7} N.\ Dorey, V.V.\ Khoze and M.P.\ Mattis, A two-instanton test of
 the exact solution of N=2 supersymmetric QCD, hep-th/9607066
\bibitem{8} H.\ Aoyama, T.\ Harano, M.\ Sato and S.\ Wada, Multi-instanton
 calculus in N=2 supersymmetric QCD, hep-th/9607076
\vfill
\end